# Ultra-soft liquid-ferrofluid interfaces

*Arvind Arun Dev*[*A,B], Thomas Hermans*[*C] and Bernard Doudin*[*A]*

A. Institut de Physique et Chimie des Matériaux de Strasbourg, UMR 7504 CNRS-UdS, 67034 Strasbourg, France
B. Laboratoire Colloïdes et Materiaux Divises, CNRS UMR 8231, Chemistry Biology & Innovation, ESPCI Paris, PSL Research University, 10 rue Vauquelin, 75005 Paris, France.
C. IMDEA Nanociencia, C/ Faraday 9, 28049 Madrid, Spain.

E-mail: arvind.dev@espci.psl.eu, thomas.hermans@imdea.org, bernard.doudin@ipcms.unistra.fr

Soft interfaces are ubiquitous in nature, governing quintessential hydrodynamics functions, like lubrication, stability and cargo transport. It is shown here how a magnetic force field at a magnetic-nonmagnetic fluid interface results in an ultra-soft interface with nonlinear elasticity and tunable viscous shear properties. The balance between magnetic pressure, viscous stress and Laplace pressure results in a deformed and stable liquid-in-liquid tube with apparent elasticity in the range 2 kPa -10 kPa, possibly extended by a proper choice of liquid properties. Such highly deformable liquid-liquid interfaces of arbitrary shape with vanishing viscous shear open doors to unique microfluidic phenomena, biomaterial flows and complex biosystems mimicking.

## 1. Introduction

Deformable interfaces govern many important hydrodynamics phenomena, for example: droplet wetting [1], lift force on a particle at the soft interface[2], suppression of flow fingering instability[3], and even laminar-to-turbulent transition at lower Reynolds numbers[4]. For soft interfaces, experiments mostly rely on elastomers since they can easily be formed into complex shapes. The elastocapillary length scale i.e., the ratio between surface tension and elasticity governs the physics of the interface. This results in phenomena like soft lubrication[5] (**Figure 1**a) with various elastohydrodynamic lubrication regimes[6], capillary bridging[7] (Figure 1b), viscous peeling of elastic sheets[8], and soft swimming[9]. The hydrodynamic mediated elastocapillary interactions result in new interfacial properties, which have been used in biomimetic system[10,11], to understand blood flow in elastic arteries[12], to access deformable cell mechanics [13], or to make fluidic analogues of electronic circuits[14–16].

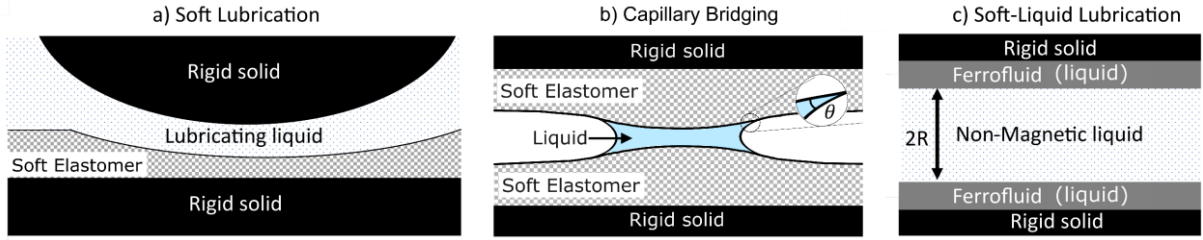

**Figure 1.** Elastocapillary and lubrication phenomena with soft interfaces. a) *Soft lubrication*[5], a balance of viscous shear and elasticity of the soft solid. The soft elastomer is deformed when there is a relative displacement between soft surface and rigid solid (top), b) *Capillary bridging* (inspired from[7]), a balance of surface tension and elasticity of a soft elastomer. A liquid element in a sufficiently soft elastometric flow channel pulls the soft elastomer using the capillary force with a contact angle $\theta$, c) "*Soft-Liquid Lubrication*", our current approach, which balances interfacial tension, viscous shear and magnetic pressure. The deformed ferrofluid topology due to pressure driven flow is discussed.

However, despite their interesting elastocapillary behavior, soft interfaces described to date still experience large viscous drag, limiting their miniaturization possibility and low-pressure deformation applications. A tunable viscous shear is essential to avoid jamming of microchannels[17], transport of concentrated materials[18], stimulation of stem cell differentiation[19], or safe transport of biomaterials through constrained geometries [20]. However, the elasticity of common elastomers is a material property, fixed at the fabrication stage and limiting its in-situ tuning. We detail here how a liquid-in-liquid flow design [21] (Figure 1c) stabilized by magnetic forces enables ultra-soft interfaces that mitigate the limitations of soft elastomers. A magnetic ferrofluid at the flow wall is used as lubricant (Figure 1c) held in place by a properly designed quadrupolar magnetic force field. The interface between the ferrofluid and a second non-magnetic liquid is highly deformable and is stabilized by competing magnetic pressure and interfacial tension. The magnetic pressure is of the order $\sim \mu_0 M H$, where $\mu_0$ is magnetic permeability, $M$ and $H$ are magnetization ($M = M(H)$) of ferrofluid and applied magnetic field respectively. The quantity $\mu_0 M H$ is analogous to elasticity. Balancing this against the interfacial tension $\sigma$, we define the magnetocapillary length scale $N_m = \sigma/\mu_0 M H$ by analogy to the elastocapillary length scale $N_e = \sigma/E$[22] with $E$ as the modulus of elasticity, governing stability criteria of soft solids[23]. Note that since $M(H)$ and $H$ can be tuned externally, $N_m$ can be tuned in-situ, consequently altering wetting behavior, like the contact angle $\theta$, see inset in Figure 1b.

## 2. Results and discussion

To elucidate and characterize this deformable "*soft-interface*" we use a magnetic forces design of nearly cylindrical symmetry that stabilizes a ferrofluid liquid envelope around an inner immiscible liquid tube geometry[21], called 'antitube'[24] (see inset of **Figure 2**a and supplementary information S1 for design principle and magnetic field calculations). We detail the fundamentals of the response of this liquid-liquid interface under dynamic conditions, measure its reversible deformation in response to pressure driven flow, and make explicit its elasticity and temporal pressure relaxation.

The antitube size and shape was measured by X-ray radiography[21,25] in 2D using a homemade setup[21] and in 3D using a RX-Solutions Easy Tom 150/160 tomographic setup (see supplementary information S2). Figure 2a shows how glycerol flow at Q = 75 µL/min leads to a conical shape with increased diameter at the glycerol inlet and decreased size at the outlet, analyzed using radiography images and an edge detector algorithm [26]. The antitube diameter at rest (no flow) is shown by a line and is measured $D_0 = 1.08 \pm 0.01$ mm. The deformation is reversible and the initial shape and size is completely recoverable upon stopping the flow (see Figure 2a and supplementary information S3). By integrating the shear stress ($\tau(r)$) over the circular cross-section, the shear force ($F$) along the antitube length can be calculated (Fig 2b), which is drastically lower (~3×) as compared to an identical solid (no-slip) wall tube (see supplementary information S4 for shear force calculation). The shear force ($F$) increases as the antitube diameter ($D$) decreases (along the length, z), following the expected increase of a corresponding solid tube behavior ($F \propto 1/D$). Note that for solid tube the shear force is fixed by the flow and geometric properties however for the same flow and geometric properties, in antitube, the shear force can be tuned by the viscosity ratio of the ferrofluid and antitube liquid[25].

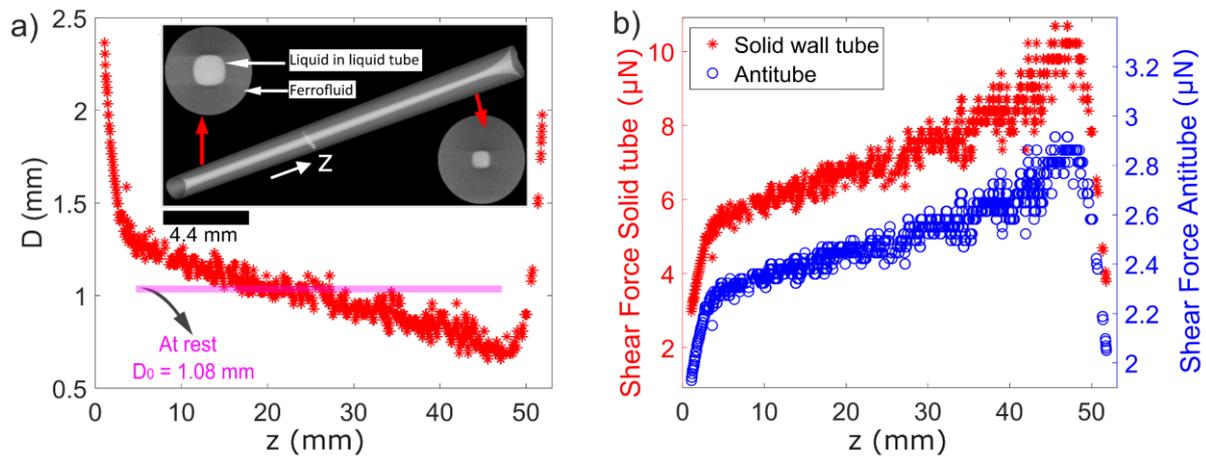

**Figure 2.** Deformable antitube where glycerol (bright) flows through APGE32 ferrofluid (darker) at $Q$ = 75 µL/min, viscosity ratio, $\eta_r = 0.65$. a) Experimentally determined diameter (D) along the flow direction (z) by X-ray tomography. The cross sections emphasize the difference between inlet (left) and outlet (right). b) Calculated shear force using the D vs z information in Fig 2a for an equivalent solid tube and antitube.

Reversible deformations of a glycerol antitube (central bright part Figure 2a inset) were measured with three different ferrofluids: APG1141, APGE32 and APG314 in order to span values of viscosity ratios ($\eta_r = \eta_{antitube}/\eta_{ferrofluid}$) 0.10, 0.65 and 4.78, respectively. For each $\eta_r$, three different no-flow diameters ($D_0$), were set by adding different amounts of ferrofluid to the cavity first filled with glycerol. For each $D_0$, three different flow rates ($Q$) in succession were applied using a syringe pump (Harvard apparatus PHD 2000) and deformations were imaged. The flow was then stopped and the temporal relaxation of pressure recorded. 3D tomography in Figure 2a inset and radiography imaging detailed in the supplementary information S3 confirms the stability and the reversibility of the observed conical shape of the antitube under flow.

**Figure 3**a-c tracks[26] the evolution of the antitube diameter along the flow direction (z) by radiography before flow (Figure 3a) and during flow (Figure 3b,c). The deformation is more pronounced with increasing $Q$ and decreasing $D_0$, due to the increased fluid shear and the lower magnetic pressure when decreasing $D_0$. This hints towards a balance between viscous shear and magnetic pressure as expected from scaling analysis (see supplementary information S4 for scaling analysis).

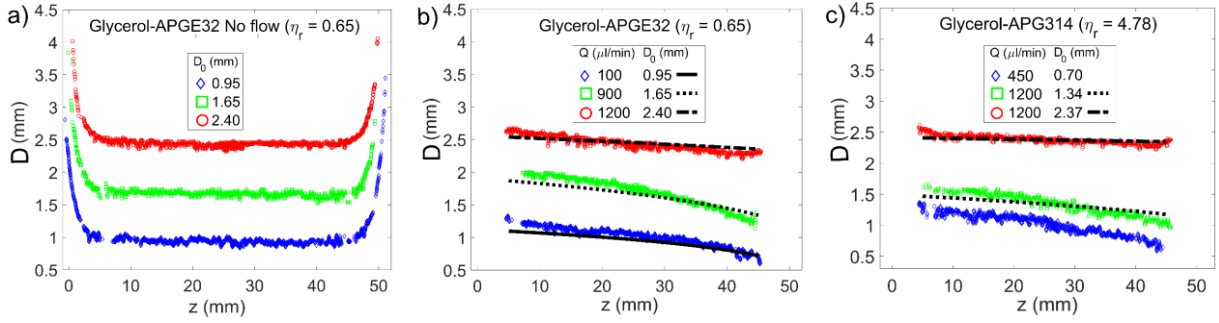

**Figure 3.** Deformation of liquid tube along its flow direction $z$. a) Diameter profile without flow for glycerol-APGE32 system, with three different $D_0$ (measured without considering inlet and outlet). b) Deformed diameter profile under flow rate $Q$. c) Deformed diameter profile for glycerol-APG314. $\eta_r$ is the viscosity ratio of the two liquids. Lines are analytical predictions from Equation (3). Increasing flow rate beyond a critical value, results in leakage (see supplementary information S3)

For quantitative insight, we consider the excess pressure[27] inside the antitube, $P = P_L + P_M = \sigma\kappa + \left(\frac{1}{2}\mu_0 M_I^2 + \int_0^{H_{Max}} \mu_0 M(H)dH\right)$, with the Laplace pressure $P_L$ resulting from the interfacial tension $\sigma$ at the liquid-liquid interface (see supplementary information S5 for interfacial tension measurements), with curvature under small slope assumption $\kappa = \frac{1}{R} - R''$, (R is the radius of the antitube and $'$ denotes the derivative with respect to the flow direction $z$). The pressure is augmented by the occurrence of Maxwell magnetic stress on the magnetic liquid $P_M$, sum of the magnetic traction at the magnetic-nonmagnetic interface $\frac{1}{2}\mu_0 M_I^2$, with $\mu_0$ the magnetic permeability, $M_I$ the magnetization at the interface that follows a cylindrical symmetry in first approximation, and the magnetic pressure due to ferrofluid magnetization, $\int_0^{H_{Max}} \mu_0 M(H)dH$ [28]. The axial gradient of the excess pressure responsible for the flow[27] is given by Equation (2)

$$\frac{dP}{dz} = \frac{\partial}{\partial z}\left(\int_0^{H_{Max}} \mu_0 M(H)dH\right) + \frac{\partial}{\partial z}\left(\frac{1}{2}\mu_0 M_I^2\right) + \frac{\partial}{\partial z}(\sigma\kappa) = -128\eta_a \frac{Q}{\pi D^4 \beta_D} \quad (2)$$

$\eta_a$ is the viscosity of the antitube liquid (here glycerol = 1.1 Pa s), $Q$ the flow rate and $\beta_D$ the drag reduction factor due to the liquid-liquid interface[25]. Since $D(z)$ has a small slope in the region excluding curved inlet and outlet, we neglect $D'''$ that scales as $D_0/L^3$, where $L$ is the length of the flow channel system (see details in supplementary information S4). This result in:

$$D' = \frac{-32\eta_a Q}{\pi D^4 \beta\left[\frac{M_r}{\pi W}A[1+B] - \frac{\sigma}{D^2}\right]} \quad (3)$$

where A and B are functions of $H$ at the interface. Equation (3) is solved numerically for the conservation of ferrofluid volume, $\int_0^L D^2 \, dz = D_0^2 L$. The resulting calculated interface profile $D(z)$ compares well with the experimental data in Figure 3b and Figure 3c. The deformation of the liquid-liquid interface under flow can be explained by the balance of magnetic pressure and viscous shear when the Laplace pressure is negligible for large D. The balance therefore results in a length scale $l_M = \left(2\eta_a QWL/\mu_0 M_r A(1+B)\beta_{D_0}\right)^{1/5}$ (see full derivation in SI S4 ).

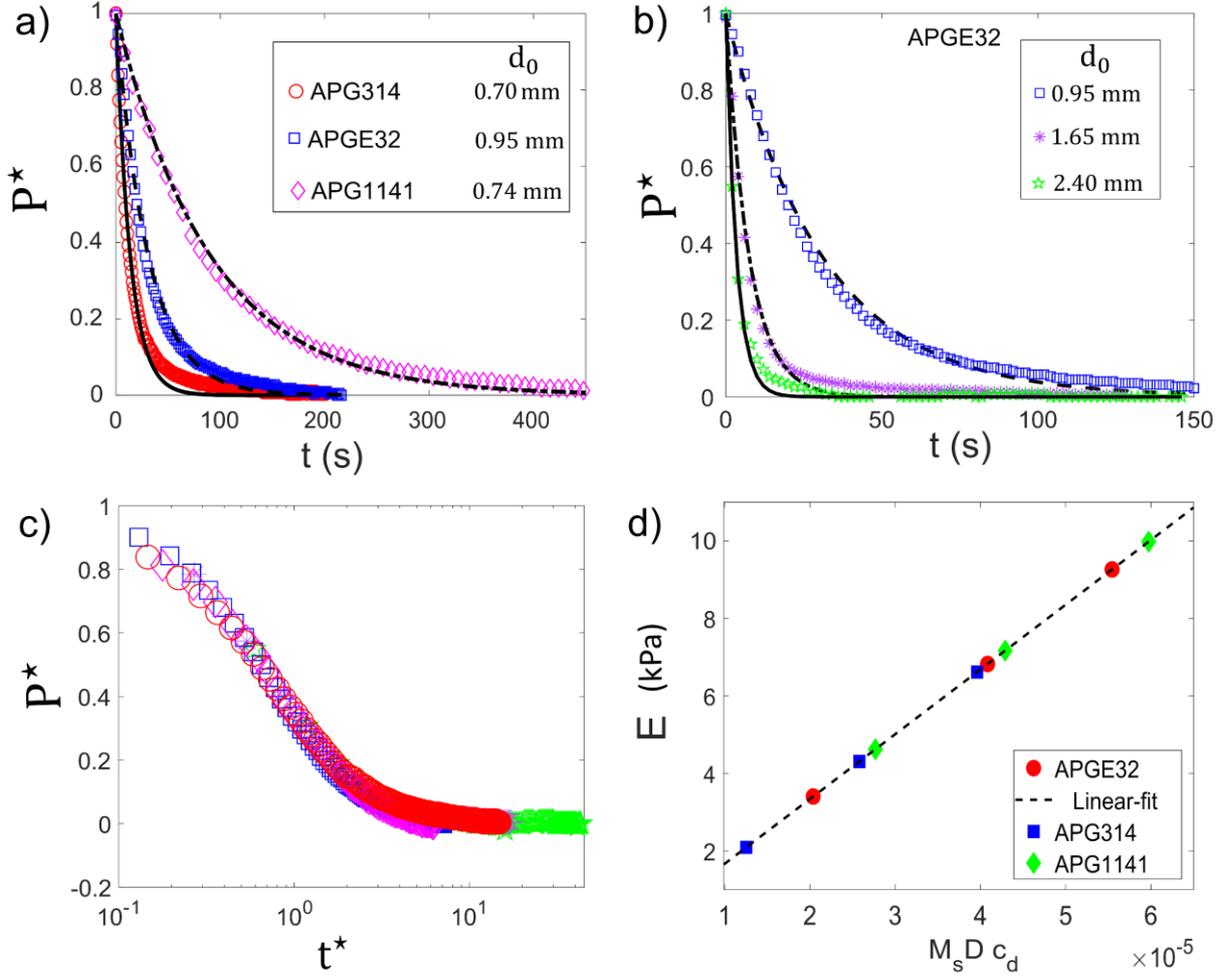

**Figure 4.** Pressure relaxation of antitubes flow, varying the diameter, the relative viscosity of the two liquids, and the flow rates. a) Pressure relaxation curve for three ferrofluids (APG314, APGE32, APG1141) , b) Relaxation curve for APGE32 at different flow diameter, c) Pressure profile in nondimensional coordinate with timescale of relaxation $t_r$ as $t^\star = t/t_r$, d) Mapped apparent elasticity of antitube for different combinations of diameter $D$ and saturation magnetization $M_s$. $P^\star$ is the non-dimensional pressure at inlet and $c_d$ is the fitting parameter to account for the non-ideal magneto-fluidic assumptions (see text for details).

The deviation of the data for largest flow rate in Figure 3c can be related to a flow regime at the onset of limits of ferrofluid confinement. Insight into elastic properties of the liquid-liquid interface can be gained by measuring the flow response under dynamic conditions. With pressure at inlet as $P_{in}(t)$, the pressure relaxation curve is obtained on stopping the flow at t=0. The pressure relaxation depends on $\eta_f$, $D$ and $M_s$, shown in **Figure 4**a for different ferrofluids, and Figure 4b for the same ferrofluid and different $D_0$ values. Relaxation is relatively slower for smaller $\eta_r$ and $D_0$ and follows $P_{in}(t) = P_{in}(0)e^{-\frac{t}{t_r}}$, where $t_r = \frac{32\pi\eta_a L^2}{E_a \beta_{D_0} D_0^2}$ is the timescale of relaxation following the literature on the cylindrical case [29], introducing the elasticity $E_a$ of the magnetic-nonmagnetic interface. Figure 4c shows that relaxation curves collapse in a reduced plot (for clarity only curves from Figure 4a and Figure 4b are shown) with $P^\star = \frac{P_{in}(t) - P_{in}(\infty)}{P_{in}(0) - P_{in}(\infty)}$ and $t^\star = t/t_r$, confirming that the experimental data is captured by $t_r$ across the experimental range. $E_a = c_d E_{m-sat}$ estimates the elasticity of the liquid-liquid interface in cylindrical architecture. $E_{m-sat} = \mu_0 MH = \frac{\mu_0 M_s M_r D}{W}$ is the value of the apparent elasticity under the hypothesis of a magnetically saturated ferrofluid (see SI S4 for the relevant discussion) and the fitted correction parameter $c_d$ ranges from 0.75-1.5 for different soft-liquid tubes and diameters. This simplified factor compensates for the deviation of the ferrofluid magnetization from saturation as well the use of an average $\beta$ factor when $\beta(D)$ slightly varies with the diameter $D$. The resulting elastic modulus is of the range 2 kPa to 10 kPa, making it an ultra-soft stable liquid-liquid interface (Figure 4d) corresponding to most soft biological tissues (thyroid, lung, breast tumor) [30]), at the lower range values (1-100 kPa)[31,32] defining soft solids. Fig 4d shows the modulus increases approximately with the diameter. When increasing the diameter (for a given solid limiting cavity and separation between source magnets), the applied field H increases, and the increased magnetic pressure makes the interface stiffer. This is a remarkable property of the system, with its nonlinear elastic properties making it analogous to biological interfaces which often exhibit strain stiffening behavior. A well-known example is blood transport vessels that exhibit a stiffness that increases with their diameter[33].

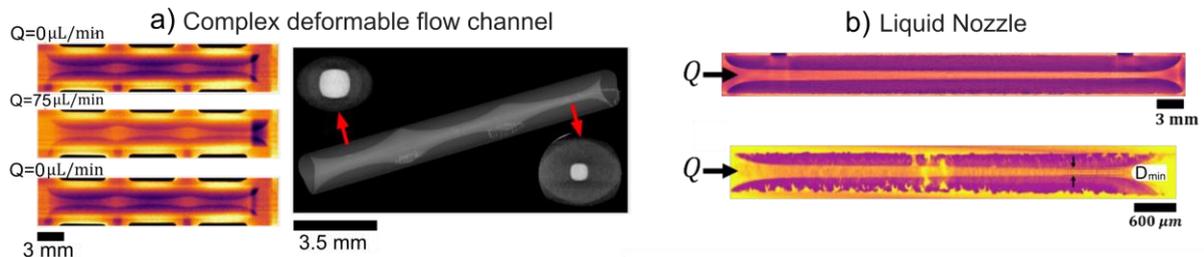

**Figure 5.** Reconfigurable microfluidics of complex shapes and flows. a) Stable oscillating (in space) shape liquid flow channel resembling RPI mode shape before flow (Q = 0 μL/min), during flow (Q = 75 μL/min) and after flow (Q = 0 μL/min), 3D X-ray tomography (greyscale) of oscillating flow channel under flow (Q = 75 μL/min), with cross sections at inlet and outlet. The permanent magnets are shown in black. b) Liquid nozzle with Glycerol-APGE32 ($\eta_r$ = 0.65), $D_{min}$ = 0.5 mm, w = 6 mm (top) and Glycerol-EMG900 ($\eta_r$ = 10), $D_{min} = 96 \pm 6\ \mu m$, w = 0.9 mm (bottom).

**Figure 5** illustrates how our low shear ultra-soft interface enables unique functional flow shapes. Figure 5a illustrates the possible 'stabilization of the instability', where magnetic forces engineered by adding three spaced magnetic quadrupoles in series are stabilize an undulating cylindrical shape (Figure 5a) resembling the Rayleigh-Plateau instability (RPI) mode shape[27,34]. Such stable liquid cylindrical structures cannot be achieved by any other reported stabilization technique. It is reversibly stable at rest (Q = 0 μL/min), confirming that there is no lower flow limit for stability. This can open doors for complex reconfigurable microfluidics design [35]. Figure 5b shows a conical liquid-in-liquid tube mimicking a nozzle shape with nozzle diameter of mm to $\mu m$ sizes with diameters reaching the microfluidic size range ($D_{min} = 96 \pm 6\ \mu m$) that can be focused (see supplementary information S6 for experimental setup), with no physics prohibiting smaller reachable diameters. Beyond viscous dominated flow ($Re < 1$), the concept can be extended to create a high throughput, soft microfluidic circuit with $Re \gg 1$ which does not shear away the ferrofluid with Reynolds numbers of several thousands (see supplementary information S6 for shearing details), even to turbulent flows for mm scale devices[36]. Such unique capability could be applied to mimic unsteady flows with deformable walls encountered in living systems (see supplementary information S6). Potentially any arbitrary shape can be created, which is an essential feature required in biomimicking and flow chemistry.

## 3. Conclusion

In conclusion, the combination of magnetic pressure, viscous stress and Laplace pressure results in a realization of stable and robust "*soft-liquid interfaces*". These interfaces can be reversibly deformed and exhibit low shear, of value than can be set by design of the viscosity of the lubricating magnetic layer. We investigate quantitatively this phenomenon in a cylindrical liquid-in-liquid flow arrangement which allows novel flow conditions, with stabilizing a liquid shape of cylindrical symmetry within a liquid environment. The lubrication effect reduces the viscous shear drastically and the balance of magnetic and viscous stress allows us to form a "*soft liquid-in-liquid nozzle*" of mm and sub-mm sizes which can find applications for example in bio-printing with minimum shear. Though a large shear system can stabilize a liquid-in-liquid flow, the magnetic encapsulation allows for stable deformed shapes with low shear characteristics and essentially without the need of continuous sheath flow since the magnetic confinement force field keeps the ferrofluid in the devices for a wide range of flow conditions. We identify a dynamic length scale $l_M$ (ratio of viscous and magnetic parameters) governing the flow induced deformation characteristics. We can therefore map elasticity to magnetic parameters, with experimental values in the range of 2-10 kPa corresponding to very soft materials. More importantly the elasticity is non-linear as it depends on the flow diameter and could be tuned in-situ. For a given permanent magnetic source (fixed $M_r = 1.2$ T), the elasticity is proportional to the saturation magnetization of the ferrofluid ($M_s$) and the geometric parameter $\frac{D}{W}$. With $\frac{D}{W} \sim 1$ and $M_s = 100$ mT (EMG 900 from ferrotech), the elasticity could be increased upto 100 kPa to cover the spectrum of soft materials [37,38] Similarly, smaller values ($< 1$ kPa) could be obtained by decreasing $\frac{D}{W}$ and $M_s$. The balance of magnetic and fluid forces thus forms a new pathway to realize liquid interface of complex architecture that can be employed in a variety of applications and opening the door to unique bio-mimicking flow conditions[39–41].

**Supporting Information**

Supporting Information is available online.

**Acknowledgements**

We acknowledge numerous fruitful discussions with Dr. P. Dunne. We thank Prof. Gauthier, Antoine Egele, and Damien Favier of Institut Charles Sadron, Strasbourg, France for the use of the EasyTom X-ray Tomography facility. This project has received funding from the European Union's Horizon 2020 research and innovation programme under the Marie Skłodowska-Curie grant agreement MAMI No 766007 and QUSTEC No. 847471. We also acknowledge the support of the University of Strasbourg Institute for Advanced Studies (USIAS) and the Fondation Jean-Marie Lehn, as well as the support from IdEx Unistra (ANR 10 IDEX 0002), SFRI STRAT'US project (ANR 20 SFRI 0012) and EUR QMAT ANR-17-EURE-0024 under the framework of the French Investments for the Future Program. Support from the Institut Universitaire de France is also gratefully acknowledged (B.D. and T.H.)

References

[1] B. Andreotti, J. H. Snoeijer, *Annu. Rev. Fluid Mech.* **2020**, *52*, 285.
[2] H. S. Davies, D. Débarre, N. El Amri, C. Verdier, R. P. Richter, L. Bureau, *Phys. Rev. Lett.* **2018**, *120*, 198001.
[3] D. Pihler-Puzović, P. Illien, M. Heil, A. Juel, *Phys. Rev. Lett.* **2012**, *108*, 074502.
[4] M. K. S. Verma, V. Kumaran, *J. Fluid Mech.* **2012**, *705*, 322.
[5] J. M. Skotheim, L. Mahadevan, *Phys. Rev. Lett.* **2004**, *92*, 245509.
[6] H. Dong, N. Moyle, H. Wu, C. Y. Khripin, C.-Y. Hui, A. Jagota, *Adv. Mater.* **2023**, *35*, 2211044.
[7] J. S. Wexler, T. M. Heard, H. A. Stone, *Phys. Rev. Lett.* **2014**, *112*, 066102.
[8] A. E. Hosoi, L. Mahadevan, *Phys. Rev. Lett.* **2004**, *93*, 137802.
[9] R. Trouilloud, T. S. Yu, A. E. Hosoi, E. Lauga, *Phys. Rev. Lett.* **2008**, *101*, 048102.
[10] M. Kim, S. Yoo, H. E. Jeong, M. K. Kwak, *Nat. Commun.* **2022**, *13*, 5181.
[11] P. M. Reis, J. Hure, S. Jung, J. W. M. Bush, C. Clanet, *Soft Matter* **2010**, *6*, 5705.
[12] Y. Ma, J. Choi, A. Hourlier-Fargette, Y. Xue, H. U. Chung, J. Y. Lee, X. Wang, Z. Xie, D. Kang, H. Wang, S. Han, S.-K. Kang, Y. Kang, X. Yu, M. J. Slepian, M. S. Raj, J. B. Model, X. Feng, R. Ghaffari, J. A. Rogers, Y. Huang, *Proc. Natl. Acad. Sci.* **2018**, *115*, 11144.
[13] L. Lanotte, J. Mauer, S. Mendez, D. A. Fedosov, J.-M. Fromental, V. Claveria, F. Nicoud, G. Gompper, M. Abkarian, *Proc. Natl. Acad. Sci.* **2016**, *113*, 13289.
[14] E. W. Lam, G. A. Cooksey, B. A. Finlayson, A. Folch, *Appl. Phys. Lett.* **2006**, *89*, 164105.
[15] F. Box, G. G. Peng, D. Pihler-Puzović, A. Juel, *Proc. Natl. Acad. Sci.* **2020**, *117*, 30228.
[16] D. C. Leslie, C. J. Easley, E. Seker, J. M. Karlinsey, M. Utz, M. R. Begley, J. P. Landers, *Nat. Phys.* **2009**, *5*, 231.
[17] E. Bertrand, J. Bibette, V. Schmitt, *Phys. Rev. E* **2002**, *66*, 060401.
[18] V. Jayaprakash, M. Costalonga, S. Dhulipala, K. K. Varanasi, *Adv. Healthc. Mater.* **2020**, *9*, 2001022.
[19] Y. Huang, J.-Y. Qian, H. Cheng, X.-M. Li, *World J. Stem Cells* **2021**, *13*, 894.
[20] A. Blaeser, D. F. Duarte Campos, U. Puster, W. Richtering, M. M. Stevens, H. Fischer, *Adv. Healthc. Mater.* **2016**, *5*, 326.
[21] P. Dunne, T. Adachi, A. A. Dev, A. Sorrenti, L. Giacchetti, A. Bonnin, C. Bourdon, P. H. Mangin, J. M. D. Coey, B. Doudin, T. M. Hermans, *Nature* **2020**, *581*, 58.
[22] B. Roman, J. Bico, *J. Phys. Condens. Matter* **2010**, *22*, 493101.
[23] S. Mora, T. Phou, J.-M. Fromental, L. M. Pismen, Y. Pomeau, *Phys. Rev. Lett.* **2010**, *105*, 214301.
[24] J. M. D. Coey, R. Aogaki, F. Byrne, P. Stamenov, *Proc. Natl. Acad. Sci.* **2009**, *106*, 8811.
[25] A. A. Dev, P. Dunne, T. M. Hermans, B. Doudin, *Langmuir* **2022**, *38*, 719.
[26] A. A. Dev, R. Dey, F. Mugele, *Soft Matter* **2019**, *15*, 9840.
[27] S. Haefner, M. Benzaquen, O. Bäumchen, T. Salez, R. Peters, J. D. McGraw, K. Jacobs, E. Raphaël, K. Dalnoki-Veress, *Nat. Commun.* **2015**, *6*, 7409.
[28] R. E. Rosensweig, *Ferrohydrodynamics*, Dover Publications, Incorporated, **2013**.
[29] G. Guyard, F. Restagno, J. D. McGraw, *Phys. Rev. Lett.* **2022**, *129*, 204501.
[30] I. Levental, P. C. Georges, P. A. Janmey, *Soft Matter* **2007**, *3*, 299.
[31] R. Pericet-Cámara, A. Best, H.-J. Butt, E. Bonaccurso, *Langmuir ACS J. Surf. Colloids* **2008**, *24*, 10565.
[32] R. W. Style, R. Boltyanskiy, Y. Che, J. S. Wettlaufer, L. A. Wilen, E. R. Dufresne, *Phys. Rev. Lett.* **2013**, *110*, 066103.
[33] X. Wang, K. Li, Y. Yuan, N. Zhang, Z. Zou, Y. Wang, S. Yan, X. Li, P. Zhao, Q. Li, *ACS Biomater. Sci. Eng.* **2024**, *10*, 3631.
[34] A. S. Utada, A. Fernandez-Nieves, H. A. Stone, D. A. Weitz, *Phys. Rev. Lett.* **2007**, *99*, 094502.

[35] J. Forth, P. Y. Kim, G. Xie, X. Liu, B. A. Helms, T. P. Russell, *Adv. Mater.* **2019**, *31*, 1806370.
[36] L. M. Stancanelli, E. Secchi, M. Holzner, *Commun. Phys.* **2024**, *7*, 1.
[37] T. Su, M. Xu, F. Lu, Q. Chang, *RSC Adv.* **2022**, *12*, 24501.
[38] S. S. Sheiko, A. V. Dobrynin, *Macromolecules* **2019**, *52*, 7531.
[39] Z. Chen, L. Fan, S. Chen, H. Zhao, Q. Zhang, Y. Qu, Y. Huang, X. Yu, D. Sun, *Adv. Healthc. Mater. n/a*, 2304532.
[40] A. Tejo-Otero, F. Fenollosa-Artés, I. Achaerandio, S. Rey-Vinolas, I. Buj-Corral, M. Á. Mateos-Timoneda, E. Engel, *Gels* **2022**, *8*, 40.
[41] Y. C. Jung, B. Bhushan, *J. Phys. Condens. Matter* **2009**, *22*, 035104.